# Network Coding in a Multicast Switch

Jay Kumar Sundararajan*, Muriel Médard*, MinJi Kim*, Atilla Eryilmaz*, Devavrat Shah*, Ralf Koetter†

*Laboratory for Information and Decision Systems,
Massachusetts Institute of Technology, Cambridge, MA 02139.
Email: {jaykumar, medard, minjikim, eryilmaz, devavrat}@mit.edu

†Coordinated Science Laboratory,
University of Illinois Urbana-Champaign, Urbana, IL 61801
Email: koetter@uiuc.edu

*Abstract*— We consider the problem of serving multicast flows in a crossbar switch. We show that linear network coding across packets of a flow can sustain traffic patterns that cannot be served if network coding were not allowed. Thus, network coding leads to a larger rate region in a multicast crossbar switch. We demonstrate a traffic pattern which requires a switch speedup if coding is not allowed, whereas, with coding the speedup requirement is eliminated completely. In addition to throughput benefits, coding simplifies the characterization of the rate region. We give a graph-theoretic characterization of the rate region with fanout splitting and intra-flow coding, in terms of the stable set polytope of the "enhanced conflict graph" of the traffic pattern. Such a formulation is not known in the case of fanout splitting without coding. We show that computing the offline schedule (*i.e.* using prior knowledge of the flow arrival rates) can be reduced to certain graph coloring problems. Finally, we propose online algorithms (*i.e.* using only the current queue occupancy information) for multicast scheduling based on our graph-theoretic formulation. In particular, we show that a maximum weighted stable set algorithm stabilizes the queues for all rates within the rate region.

## I. INTRODUCTION

The input-queued crossbar switch has been studied well, especially in the context of unicast traffic. It is known that 100% throughput can be achieved [1], in the sense that as long as no input or output is over-subscribed, traffic can be supported without causing the queues to grow unboundedly. This is accomplished by the maximum weighted bipartite matching algorithm, with queue lengths as weights. Several simplifications of this algorithm have also been investigated ([2], [3], [4]). If the arrival rates are known in advance (or if they are estimated in real time), we can use the capacity decomposition approach of the Birkhoff-von Neumann switch [5], [6]. Reference [7] contains a good summary of the unicast switching literature.

The extension of the problem to multicast flows is intrinsically more difficult. Early approaches used the copy strategy – make copies of the cells[1] in a separate stage before the switching fabric, and then treat them like unicast flows [8]. However, this approach reduced the bandwidth available to other traffic in the switch. It became clear that the *intrinsic multicast capability*[2] of the switching fabric must be utilized.

Prabhakar *et al.* [9] studied the tradeoff between throughput and complexity of two heuristic algorithms, under a fairness constraint. The hardness of the multicast scheduling problem was proved by Andrews *et al.* [10]. They also showed the hardness of the problem of integrating unicasts along with the multicasts.

Marsan *et al.* [11] gave a characterization of the rate region achievable in a multicast switch with fanout splitting[3], and also defined the optimal scheduling policy. Interestingly, this work proved that unlike in the unicast case, 100% throughput cannot be achieved for multicast flows in an input-queued switch. In fact, the minimum speedup needed to achieve 100% throughput grows unboundedly with the switch size.

Our paper studies the same problem as [11], with the following modification. The inputs are allowed to send linear combinations of cells waiting in the queues, *i.e.*, they are allowed to perform linear network coding. We show that this modification enables a number of interesting and non-trivial benefits as well as insight. The main contributions of this paper are:

1) We prove that linear network coding increases the achievable rate region of the switch.
2) We provide a simple graph-theoretic characterization of the rate region with coding, which in turn leads to more insight on the problem.
3) We provide offline and online algorithms to achieve this rate region while stabilizing the queues.

Reference [12] gave one example to show the benefit of network coding in a multicast switch, which we will revisit in Section II. The paper also gave an outer bound on the rate region with fanout splitting and intra-flow coding. In this paper, we prove that this bound is indeed achievable.

In the special case when fanout splitting is not allowed for any flow, [13] showed that the rate region is the stable set polytope of a suitably defined "conflict graph". A similar graph-theoretic formulation was used by Caramanis *et al.* in [14] in the context of unicast traffic in Banyan networks. References [13] and [15] showed that, if the flow rates are known in advance, then a rate decomposition based approach can be used to compute the schedule, in a manner similar to the Birkhoff-von Neumann unicast switch [5]. They also showed that such rate decomposition reduces to fractional weighted coloring of the conflict graph. In this paper, we extend these graph-theoretic connections and insights to the case when fanout splitting and coding are both allowed.

---

[1] Packets arriving at the switch are split into fixed-size units called *cells* which are reassembled into packets at the output.

[2] The ability to transfer simultaneously, a cell to multiple outputs using simultaneous switching paths

[3] Fanout splitting is the ability to serve partially, a multicast cell to only a subset of its destined outputs, and complete the service in subsequent timeslots.

Note that, for the case of fanout splitting without coding, [11] gave a characterization of the rate region as the convex hull of certain modified departure vectors. However, a graph-theoretic formulation of the same is not known. On the other hand, for the case with coding, our graph-theoretic formulation helps us understand the effect of the traffic pattern on the throughput. We transform any given traffic pattern into a conflict graph, and the properties of this graph can be used to derive insight on what kind of traffic patterns are "hard" in terms of computing the schedule, and in terms of achieving 100% throughput.

The rest of the paper is organized as follows. Section II gives an example of a traffic pattern that cannot be achieved with fanout splitting alone, but can be achieved when network coding is allowed. Section III introduces the concept of "enhanced conflict graph" and gives a graph-theoretic formulation of the problem. It also states our main theorems on the rate region and the computation of an offline schedule using a rate decomposition based approach. Section IV uses these theorems to quantify the benefits due to coding, in two situations – the $2 \times 3$ switch, and the $2 \times N$ switch, with a traffic pattern that generalizes the example of Section II. In the latter case, we prove that a speedup of around 1.5 is needed without coding, as opposed to no speedup if coding is allowed. Section V addresses the problem of finding online algorithms for optimal scheduling of a multicast switch, when fanout splitting and coding are allowed. We propose a maximum weighted stable set algorithm and show that it is optimal. Finally, in Section VI, we summarize the contributions of this paper, and discuss potential avenues for future work.

## II. NETWORK CODING IMPROVES THROUGHPUT – AN EXAMPLE

Consider the traffic pattern $T$ shown in Fig. 1. This is a $2 \times 3$ switch, with 4 flows[4] – one multicast flow from input 1 to all 3 outputs, and 3 unicast flows from input 2 to outputs 1, 2 and 3 respectively. The rates of the 4 flows are set at $\frac{2}{3}$, $\frac{1}{3}$, $\frac{1}{3}$ and $\frac{1}{3}$ respectively (normalized with respect to the arrival rates).

This traffic pattern cannot be achieved with fanout splitting. To show this, we note that at all times in the schedule, one of the unicasts from input 2 has to be served, since it is a saturated input (*i.e.* the total inflow is 1). Therefore, in any time-slot, one output is blocked and a multicast packet can, at best, be sent to the other 2 outputs. This means, it will take at least 2 slots to complete the service of each packet, and hence, a rate of more than $\frac{1}{2}$ is not achievable. Since the required rate is $\frac{2}{3}$, fanout splitting cannot achieve this traffic pattern.

However, it can be served if intra-flow network coding and fanout splitting are allowed. In fact, a code over the binary field is sufficient. The network code involves coding at input 1, over packets only from the multicast flow. Input 1 codes over blocks of 2 packets, and sends them over 3 time-slots. Consider a block of two packets $\{P_1, P_2\}$ from the multicast flow. Table I gives the code and also specifies to which outputs the coded packet is sent. The $\oplus$ sign indicates that the packets are XORed bitwise and sent. It can be verified that this code enables each of the three destinations to decode both packets in the block, at the end of 3 time-slots.

[4]The formal definition of a flow, and the switch assumptions are given in Section III.

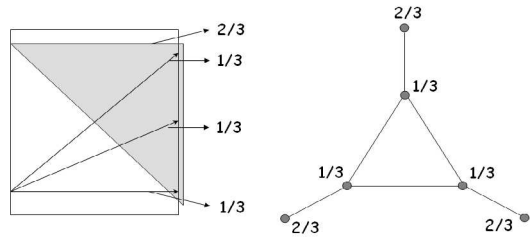

Fig. 1. The example traffic pattern showing the benefit of network coding, along with its enhanced conflict graph

| Time-Slot | Code | Outputs |
|---|---|---|
| 1 | $P_1$ | 1,2 |
| 2 | $P_2$ | 2,3 |
| 3 | $P_1 \oplus P_2$ | 3,1 |

TABLE I
THE NETWORK CODE USED BY INPUT 1

As for the unicast flows, they are also served in these time-slots, in parallel. Note that, input 1 talks to only 2 outputs at any given time (column 3 of the table). Thus, input 2 can send a unicast packet to the third unoccupied output and a rate of $\frac{1}{3}$ is achieved for each of the unicasts. In other words, the given code satisfies all the rate requirements of the example.

## III. THE RATE REGION

In this section, we present our main theorem which uses a graph-theoretic formulation to characterize the rate region with fanout splitting and coding. We begin with some definitions and a note on the queuing assumptions in the switch.

*Definition 1 (**Flow**):* A flow is a stream of packets that have a common source and destination set. It is represented by a 2-tuple $(i, J)$ consisting of the input $i$ and a subset $J$ of outputs corresponding to the destination set of the multicast stream.

*Definition 2 (**Sub-flow**):* A sub-flow is a 3-tuple $(i, J, j)$ consisting of an input $i$, a subset of outputs $J$ and one output $j$ from that set, *i.e.* $j \in J$.

A sub-flow $(i, J, j)$ is said to belong to the flow $(i, J)$. A multicast flow can be thought to consist of several sub-flows, each going to a different output.

*Queuing assumptions:* The inputs are assumed to have a separate buffer for every flow. The buffers are not restricted to be FIFO. In fact, we assume that, in any time-slot, it is possible to compute a linear combination of all packets within a buffer and send it out. However, there is no coding across contents of different buffers. The usual switch constraints are assumed to hold – an input may not send different packets to different outputs simultaneously, and an output may receive a packet from only one input at a time. For the case when coding is not allowed, we assume the multicast virtual output queue architecture with re-enqueuing, as described in [11].

Since fanout splitting is allowed, sub-flows belonging to the same flow do not conflict with each other, in the sense that any subset of them may be served simultaneously. Owing to the switch constraints, an input cannot send different information to different outputs at the same time. Moreover, we do not allow coding across packets of different flows. Hence, at any point of time, sub-flows belonging to different flows may not be served together. Therefore, any sub-flow conflicts with sub-flows of other flows at the same input. It also conflicts with sub-flows from other inputs, that are destined to its output. These conflicts are captured by the enhanced conflict graph [5].

*Definition 3 (**Enhanced Conflict Graph**):* Given a traffic pattern, the enhanced conflict graph $G = (V, E)$ is an undirected graph defined as follows:

*Vertices*: The graph contains one vertex for every sub-flow of every flow.

*Edges*: Each sub-flow[6] is connected to all sub-flows belonging to other flows at the same input. In addition, each sub-flow is also connected to all sub-flows that have the same output.

The above definition implies that the set of sub-flows that are served simultaneously in any valid configuration of the switch, must be a stable set in the enhanced conflict graph.

*Definition 4 (**Enhanced Rate Vector**):* Let $\mathbf{r} \in \mathbb{R}^f$ be the rate vector of a traffic pattern that has $f$ flows. Suppose the total number of sub-flows in the pattern (*i.e.* the sum of all the fanout sizes) is $s$. The enhanced rate vector $\mathbf{e}(\mathbf{r}) \in \mathbb{R}^s$ corresponding to $\mathbf{r}$ is defined as:

$$e_{(i,J,j)}(\mathbf{r}) = r_{(i,J)}, \text{ for all } j \in J.$$

We use the enhanced rate vector as weights for vertices of the enhanced conflict graph.

For example, Fig. 1 also shows the enhanced conflict graph for the traffic pattern on the left. Note that, the weight associated with the sub-flows of a particular multicast flow are all equal to the rate of that flow.

*Definition 5 (**Innovative Packet**):* A packet transmitted from an input to an output is said to be *innovative* if it conveys previously unknown information to an output. For linear network coding, this means that the vector of coefficients used in the linear combination while computing the packet, is linearly independent of coefficient vectors of all packets received previously by the output.

### A. The rate region with fanout splitting and intra-flow coding

By the term rate region, we mean the set of average arrival rate vectors for which there exists a schedule that serves the flows without causing the queues to build up indefinitely with time.

*Theorem 1:* The rate region with fanout splitting and intra-flow linear network coding, denoted $\Gamma$, is given by the set of all rate vectors $\mathbf{r}$ such that, the enhanced rate vector $\mathbf{e}(\mathbf{r})$ is in the stable set polytope of the enhanced conflict graph.

*Proof:*

**Achievability:** Suppose $\mathbf{e}(\mathbf{r})$ is in the stable set polytope of the enhanced conflict graph, then we can express $\mathbf{e}(\mathbf{r})$ as a convex combination of the incidence vectors of stable sets of the graph. Assuming the coefficients of the convex combination to be rational numbers, we can construct a frame-based schedule by appropriate time-sharing among the different switch configurations represented by the stable sets. In each time-slot, the stable set specifies which sub-flows are to be served. Let the length of such a frame be $F$.

This schedule has the property that every input is connected to every output for enough fraction of time, so as to satisfy the demand of each sub-flow between them. More specifically, consider one particular flow $f$ of rate $r_0$. Assume $r_0$ is rational and $F$ is large enough so that $r_0 F$ is integer and $r_0 F$ packets are served in one frame. Then, the schedule guarantees that during the course of the frame, each output in the fanout of flow $f$ receives $r_0 F$ transmissions from the input.

Thus, to prove achievability, we only need to ensure that none of the transmission opportunities is wasted. In every time-slot, each transmitted packet must be innovative to all the outputs it reaches. It is possible that there is no single packet that will be innovative to all the connected outputs. For instance, each of the outputs may have all but one packet from a common set, and the missing packet may be different for each output. This is where network coding is required. A coded packet can simultaneously satisfy all the outputs in such a situation.

We use a maximum distance separable (MDS) code [16] to prove the achievability. Let $T$ be the total number of slots in one frame, in which flow $f$ is served to any of its outputs. Owing to fanout splitting, $T$ is in general more than $r_0 F$. In the coding scheme we propose, the input uses a $(T, r_0 F)$ MDS code and computes the codeword treating the $r_0 F$ packets as symbols of the information word. Then, at each of the $T$ transmission opportunities for flow $f$, the input transmits a new symbol from the MDS codeword that it computed.

The key property of an MDS code that we use here is that an $(n, k)$ MDS code can correct upto $(n - k)$ erasures, each of which may occur in any position of the codeword. Hence, using any set of $k$ codeword symbols one can retrieve all the information. In our case, since each output in the fanout of $f$ is guaranteed to receive $r_0 F$ codeword symbols, it can retrieve the entire transmitted information. (Note: The schedule and the code are computed offline, and are known to all inputs and outputs.)

**Converse:** The proof of the converse was given in [12], and is summarized here for completeness. Let $\mathbf{r}$ be any achievable rate vector. Let $\mathbf{e}(\mathbf{r})$ be the corresponding enhanced rate vector. Since $\mathbf{r}$ is achievable, there is a schedule of switch configurations and associated codes for each time-slot in the schedule such that, every sub-flow receives innovative packets for enough fraction of time so as to meet its rate requirement. Based on this achieving schedule, form a 0-1 indicator vector in each time-slot, with one entry for each sub-flow such that, this vector has a 1 for those sub-flows for which an innovative packet is conveyed by the code in that time-slot. Then **e(r)** is the time average of such indicator vectors over all the time-slots. But then, each indicator vector has to be the incidence vector of some stable set of the enhanced conflict graph due to the switch constraints. Thus, any achievable enhanced rate vector can be expressed as a convex combination of stable sets of the enhanced conflict graph, and this proves the converse. ∎

---

[5]The term "enhanced conflict graph" is used to distinguish it from the term "conflict graph" that was used in [13], for the case of no-fanout-splitting.

[6]In the rest of this definition, by sub-flow, we mean the vertex representing this sub-flow.

Since we view the packets as elements of a finite field while computing the code, the field size required for the code is a parameter of interest. If the field is too large, then we may need more than one packet to represent a single field element, which makes the implementation more difficult. On the other hand, in order to ensure that every transmission leads to an innovative packet being conveyed to all the recipients, we must operate over a large enough field. The following discussion indicates that for reasonable assumptions on the switch size and the packet size, the field size required will be such that a field element will indeed fit within one packet.

The proof above uses an MDS code to show achievability. In general, for an $(n, k)$ MDS code to exist, we need to work over a large field size, comparable to $n$. This means the field size could depend on the length of the schedule, which is not desirable. However, using the results of [17] and [18], one can show that there exist other codes which are defined over a field as big as only the fanout size of the flows, while still ensuring that every transmission is useful to all recipients.

*Proposition 3.1:* A field size equal to the fanout size is sufficient to ensure that every transmission is innovative to all outputs, in the proof of Theorem 1.

*Proof:* We use the same notation as in the earlier proof. Consider a network with three layers of nodes. The first layer has a single node – the source. The second layer nodes correspond to the time-slots in the frame in which flow $f$ is being served. Thus, there are $T$ such nodes. In the third layer, there is one node corresponding to each output in the fanout of flow $f$. The source node is connected to all nodes in the second layer. A node in the second layer is connected to those nodes of the third layer which are served in the corresponding time-slot. All links have unit capacity. Consider the single source multicast problem with network coding, from the source node to all nodes of the third layer. Since the schedule guarantees that every output receives $r_0 F$ transmissions, this means the min-cut of this network is $r_0 F$. Therefore, using the results of [17] and [18], $r_0 F$ packets can be transmitted to each output using network coding, and the field size required is equal to the number of destinations, which in our case is the size of the fanout. The network coding solution to this new network naturally leads to the code for the switch, namely that, in the $i^{th}$ time-slot, the switch input should use the same linear combination that the source transmitted to the $i^{th}$ node of the second layer in the network. ∎

*B. Admissible region vs. rate region*

For a general graph, a complete characterization of the stable set polytope in terms of linear inequalities is unknown. Several families of necessary conditions are known. One example is the clique inequalities, which say that the total weight on the vertices of a maximal clique[7] must not exceed 1. In terms of the switch, we can show that the maximal cliques of the enhanced conflict graph correspond to flows either from the same input or to the same output. Thus, the clique inequalities imply that no input nor any output may be overloaded. This is also called the *admissibility conditions*. The polytope described by these conditions along with non-negativity constraints is called the *admissible region*.

It is known that the non-negativity and clique inequalities suffice in describing the polytope, if and only if, the graph is perfect [19]. This leads to the following corollary.

*Corollary 1: For a given traffic pattern, the entire admissible region is achievable with fanout splitting and intra-flow linear network coding, if and only if the enhanced conflict graph is perfect.*

In a more general case, the admissible region is a strict superset of the rate region. This implies that it is not possible to achieve 100% throughput even with fanout splitting and coding. The formulation presented here gives us insight on what kind of traffic patterns will lead to a reduced throughput.

Note that, for the case with fanout splitting, but no network coding, Marsan *et al.* showed in [11] that 100% throughput cannot be achieved. However, in that case, the rate region is in terms of the convex hull of all possible departure vectors, which do not have a neat graph-theoretic characterization in general. Thus, allowing network coding leads to a more insightful description of the rate region, and enables the use of graph-theoretic tools.

*C. Rate decomposition approach to compute the schedule*

In this subsection, we address the problem of the Birkhoff-von Neumann like rate decomposition approach for offline computation of the schedule, given the rates of the flows, in a manner similar to [5] and [6]. The following corollary gives a graph-theoretic interpretation of this approach.

*Definition 6 (**Fractional Weighted Coloring Problem**):* Given a graph $G$ and a weight $w_v \in \mathbb{R}^+$ for each vertex, minimize $\sum_{i=1}^{k} \lambda_i$ $(\lambda_i \in \mathbb{R}^+, \forall i)$ such that there exist stable sets $\{S_i\}$ of $G$ with $\sum_{i=1}^{k} \lambda_i \chi^{S_i} = \mathbf{w}$, where $\mathbf{w}$ is the given weight vector, and $\chi^S$ denotes the incidence vector of the stable set $S$. The optimum value of the minimization problem is called the *fractional weighted chromatic number*.

We interpret the weights to correspond to the flow rates, and the coefficients $\lambda_i$ to be the fractions of time in the schedule. Essentially, if the fractional weighted chromatic number is less than 1, then the optimal solution expresses the weight vector as a convex combination of stable sets, which in turn leads to a switch schedule. This leads to the following corollary.

*Corollary 2: The problem of computing the offline switch schedule for a multicast traffic pattern when fanout splitting*

---

[7]A clique is a set of vertices all of which are connected to each other.

and intra-flow linear network coding are allowed, is equivalent to the problem of fractional weighted coloring of the enhanced conflict graph, with the enhanced rate vector used as vertex weights.

Given the set of rates of the various flows in a multicast switch, the switch schedule can be obtained as above. This ensures that each input gets to talk to each output for enough fraction time about each flow. To make sure that every transmission opportunity is used to convey a new degree of freedom, we need to use an appropriate code. One way to do this is the MDS code idea described in the proof of Theorem 1. Alternatively, to obtain a code using a smaller field size, one can use the ideas in Proposition 3.1, where a multicast network code construction is used on a new network that represents the transmission schedule. The switch schedule and the network code which ensures that every transmission conveys an innovative packet, together give a complete specification of a frame-based scheme that achieves the entire rate region.

### D. The effect of speedup

*Definition 7 (Speedup):* A switch is said to have a speedup $s$ if the switching fabric can transfer packets at a rate $s$ times the incoming and outgoing line rate of the switch.

If we define a time-slot to be the reciprocal of the line rate, then this means the switching fabric can go through $s$ configurations within one time-slot. Note that this requires output queuing.

So far we have considered the case where $s = 1$. It is easy to see that a rate vector $\mathbf{r}$ is achievable with speedup $s$ if and only if it is admissible and $\frac{1}{s}\mathbf{r}$ is within the rate region corresponding to a speedup of 1.

Now, if the fractional weighted chromatic number $c$ (defined above) for a given rate vector exceeds 1, then such a rate vector cannot be achieved, since it is not within the stable set polytope. However, if we allow a speedup equal to $c$, then the rate can be achieved. This is because the speedup essentially scales down the rate vector by a factor of $c$, and this in turn scales down the optimum value of the minimization by the same factor. Hence, the new rate vector is inside the rate region. This gives an interesting physical interpretation for the fractional weighted chromatic number corresponding to a given rate vector, which is summarized in the following theorem.

*Theorem 2:* The minimum speedup needed to achieve a given rate vector with fanout splitting and coding, is the fractional weighted chromatic number of the enhanced conflict graph, with the enhanced rate vector used as vertex weights.

## IV. EXAMPLES AND SIMULATION

Network coding gives a benefit in the rate region, even if we use only linear intra-flow coding. In this section, we apply the formulation described in the previous section, to quantify the benefits in a $2 \times 3$ switch with arbitrary traffic, and in a $2 \times N$ switch, for a special traffic pattern.

| Polytope | Volume | Normalized Volume | Speedup to achieve $P_{adm}$ |
|---|---|---|---|
| $P_{adm}$ | $4.921 \times 10^{-9}$ | 1 | 1 |
| $P_{intra}$ | $4.686 \times 10^{-9}$ | 0.952 | 1.25 |
| $P_{fs}$ | $4.613 \times 10^{-9}$ | 0.937 | 1.25 |
| $P_{nofs}$ | $2.260 \times 10^{-9}$ | 0.460 | 1.67 |

TABLE II

A COMPARISON OF THE FOUR SCHEMES

### A. $2 \times 3$ switch

In a $2 \times 3$ switch, there are 14 possible flows - three unicasts, three two-casts, and one broadcast from each of two inputs. Thus, the rate region is a 14-dimensional polytope. We numerically computed the facets of the stable set polytope of the enhanced conflict graph corresponding to this traffic pattern. Then, we used this to obtain and study the rate region of the $2 \times 3$ switch.

The methodology we used was to first list out all stable sets of the enhanced conflict graph. These are the extreme points of the rate region. We used a package for MATLAB known as the multi-parametric toolbox [20] to convert the extreme point representation to a representation using linear inequalities, which was then used for the speedup computations. To compute the volume of the polytopes, we also used a software known as Vinci [21]. The rate region of the case with fanout splitting but no coding was obtained using the characterization given in [11].

The two rate regions were compared in terms of the volume of the polytope and the minimum speedup needed to achieve the entire admissible region. The results are summarized in Table IV-A. Here, $P_{adm}$ refers to the admissible region, $P_{intra}$ refers to the rate region with linear intra-flow network coding and fanout splitting, $P_{fs}$ is the case with only fanout splitting, and $P_{nofs}$ is the rate region when fanout splitting is not allowed. The corresponding values for the case with no fanout splitting, are also shown. These results were obtained using the graph-theoretic formulation that was obtained in [13] and [15].

The results indicate that there is a marginal improvement in throughput due to coding. There is another way to compare the two schemes. We know that network coding enlarges the rate region. The same region can also be achieved without coding, if we allowed a speedup. The amount of speedup needed for this to happen is a measure of the gain due to coding. This value is 1.1667 for the $2 \times 3$ case. Thus, for certain traffic patterns, coding can do away with the need for speedup, even in a $2 \times 3$ switch.

### B. Example in a $2 \times N$ switch

We now study a special traffic pattern in a $2 \times N$ switch, where the benefit of network coding is pronounced. Consider a $2 \times N$ switch, with the following traffic pattern: at input 1, there is one broadcast flow going to all outputs, having a rate $r_0$; at input 2, there are $N$ unicasts, one to each output – the rate of the unicast going to output $j$ is $r_j$, for $j = 1, 2, \ldots, N$. See Fig. 2.

The rate region of this pattern with fanout splitting but no

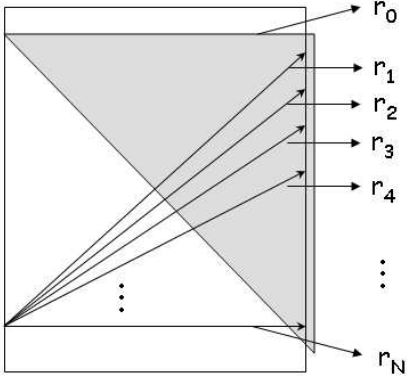

Fig. 2. A special traffic pattern which demonstrates the benefit of coding

coding, can be shown to be:

$$r_i \geq 0 \quad \text{for } i = 1, 2, \ldots N \quad (1)$$
$$\sum_{i=1}^{N} r_i \leq 1 \quad (2)$$
$$r_0 + r_i \leq 1 \quad \text{for } i = 1, 2, \ldots N \quad (3)$$
$$2r_0 + \sum_{i=1}^{N} r_i \leq 2 \quad (4)$$

It is easy to verify that these are necessary conditions. It turns out that they are sufficient as well, but we skip the proof of this fact for want of space.

Now, consider the enhanced conflict graph for this pattern.

*Theorem 3:* The enhanced conflict graph for the special pattern is a perfect graph.

*Proof:* It consists of a set of $N$ sub-flows one for each unicast, and another set of $N$ sub-flows for the broadcast. The unicast sub-flows form a clique, while the broadcast sub-flows form a stable set. Thus, the set of vertices can be partitioned into two parts, which induce a clique and a stable set respectively. This means the graph is a "split graph", which is known to be perfect [19]. ∎

The next corollary follows immediately from Section III-B.

*Corollary 3:* For the special traffic pattern, the entire admissible region is achievable if fanout splitting and linear intra-flow coding are allowed.

This means that the following rate vector is achievable: $r_0 = \left(1 - \frac{1}{N}\right)$; $r_j = \frac{1}{N}$ for all other $j$.

This fact can be verified by observing that the traffic pattern is a generalization of the example given in Section II, and that, the same single parity check code can be generalized to $N$ bits, and used here, to achieve the traffic pattern.

But, this rate vector does not lie within the rate region for fanout splitting without coding because it violates the inequality given in Eqn. 4. The left hand side evaluates to $\left(3 - \frac{2}{N}\right)$, while the right hand side is only 2. Hence, the smallest scaling factor such that the scaled rate vector lies inside the rate region is $\left(1.5 - \frac{1}{N}\right)$. In other words, we have demonstrated a traffic pattern which can be served with no speedup if network coding is allowed, but needs a speedup of $\left(1.5 - \frac{1}{N}\right)$, if coding is not allowed.

The two examples show that the throughput benefit due to coding depends on the traffic pattern in the switch. For instance, in the $2 \times 3$ case, for some traffic patterns, both fanout splitting and coding are equally far away from the admissibility limit, which is why both need the same speedup of 1.25 to cover the entire admissible region. However, for some other traffic patterns, coding needs no speedup, while fanout splitting may need a non-trivial speedup.

We expect that the gains observed will be more pronounced in larger switches, as the number of traffic patterns where coding gives a benefit will be more. However, there will still be patterns for which there is no gain due to coding. The more important gain of coding is the simpler characterization of the rate region and the insight gained from the graph-theoretic formulation.

## V. MAXIMUM WEIGHTED STABLE SET ALGORITHM FOR ONLINE SCHEDULING

Suppose the rates of the various flows are unknown, and scheduling has to be done online, using only the queue occupancy information. Analogous to the maximum weighted matching algorithm for unicast, we show that a maximum weighted stable set algorithm on the enhanced conflict graph achieves the entire rate region for multicast when fanout splitting is allowed along with network coding. In this section, we assume that the arrivals to each flow are *i.i.d.* and independent across flows.

*Lemma 1:* Let $\mathcal{V}$ be a vector space with dimension $n$ over a field of size $q$, and let $\mathcal{V}_1, \mathcal{V}_2, \ldots \mathcal{V}_k$, be subspaces of $\mathcal{V}$, of dimensions $n_1, n_2, \ldots, n_k$ respectively. Suppose that $n > n_i$ for all $i = 1, 2, \ldots, k$. Then, there exists a vector that is in $\mathcal{V}$ but is not in any of the $\mathcal{V}_i$'s, if $q > k$.

*Proof:* The total number of vectors in $\mathcal{V}$ is $q^n$. The number of vectors in $\mathcal{V}_i$ is $q^{n_i}$. Hence, the number of vectors in $\cup_{i=1}^{k} \mathcal{V}_i$ is at most $\sum_{i=1}^{k} q^{n_i}$. Now,

$$\sum_{i=1}^{k} q^{n_i} \leq k q^{n_{max}} \leq k q^{n-1} < q^n$$

where, $n_{max}$ is $\max_i n_i$, which is at most $(n-1)$.

Thus, there are more vectors in $\mathcal{V}$ than in the union of all the $\mathcal{V}_i$'s. This completes the proof. ∎

We introduce variables of the form $x_{iJj}(t)$, for every sub-flow $(i, J, j)$ in every time-slot $t$, such that, $x_{iJj}(t)$ represents the difference between the total number of packets of flow $(i, J)$ that have arrived until time $t$, and the number of innovative packets transmitted from input $i$ to output $j$, for flow $(i, J)$ until time $t$. Thus, $x_{iJj}(t)$ is a measure of the backlog for sub-flow $(i, J, j)$ in terms of the degrees of freedom.

*Theorem 4:* The maximum weighted stable set (MWSS) algorithm given above stabilizes the vector $\mathbf{x}$ in the mean, provided the rate vector is inside $\Gamma$ (the rate region given in Theorem 1).

*Proof:* First we need to show that the second step in the maximum weight stable set algorithm is feasible. Now, $x_{iJj}$ gives the difference in the number of dimensions that the input knows and the number of dimensions that output $j$ knows.

*Algorithm:* Max Weighted Stable Set (MWSS)

1. Using $x_{iJj}(t)$ as the weight for the vertex corresponding to the sub-flow $(i, J, j)$, compute the maximum weighted stable set in the enhanced conflict graph. This specifies the set of sub-flows that will be served in the current time-slot. If $x_{iJj}$ is 0 for any sub-flow, it is dropped from the stable set.
2. For every flow in the chosen set, compute a linear combination of all packets received for that flow until time $t$, such that, the linear combination is an innovative packet for all the chosen outputs of that flow.
3. Transfer the computed linear combination to the outputs of the sub-flows chosen in the stable set in step 1, and update $x_{iJj}(t)$ accordingly. Go back to step 1.

Hence, if $x_{iJj}$ is positive for a set of outputs, then Lemma 1 guarantees that there exists a linear combination of the packets of flow $(i, J)$ that is innovative to all those outputs, for a field size that is larger than the number of outputs involved. Such a combination is chosen in step 2. The network code thus ensures that every transmission opportunity is used to convey a new degree of freedom.

The rest of the proof is essentially an application of the results of [22] and [23] for the case of parallel queues. The sub-flows can be viewed as *virtual queues*. Arrivals to a sub-flow "queue" are defined to occur whenever an arrival occurs to the corresponding flow in the switch. A departure from the queue is defined to occur when a new degree of freedom is conveyed for the sub-flow. Eligible activation vectors of the queues therefore correspond to conflict-free sets of sub-flows, or in other words, stable sets in the enhanced conflict graph.

Under these definitions, the only difference between this situation and the situation assumed in [22] is that the arrivals to different queues are assumed to be independent of each other in [22], whereas in our case, arrivals to sub-flows of the same flow always occur simultaneously. However, this lack of independence across arrival processes does not affect the results of [22], essentially because of the linearity of expectation of dependent random variables. Stability in the mean still holds, as long as other assumptions such as the ergodicity of the arrival processes and the finiteness of their second moment hold. Thus, the maximum weighted stable set algorithm stabilizes the occupancy of the virtual queues, as long as their arrival rates are inside the convex hull of the eligible activation vectors, *i.e.* the stable set polytope.

But then, the above definitions of arrival and departure from the virtual queues imply that the occupancy of queue $(i, J, j)$ at any time-slot is precisely $x_{iJj}(t)$. Thus, this proves that the **x** vector is stabilized in the mean, as long as the arrival rate vector is within $\Gamma$. Thus, for all choices of $(i, J, j)$, $\lim_{t \to \infty} E[x_{iJj}(t)] < \infty$. ∎

*Finite Horizon MWSS Algorithm*

The above theorem shows, that in terms of the number of degrees of freedom, the MWSS algorithm ensures stability. However, the algorithm has no provision for packets to depart from the buffer, since all packets that have arrived until time $t$ can potentially be used for coding at time $t$. To show that the buffers can also be stabilized, we modify the algorithm into a batching scheme, which we call the *finite horizon MWSS algorithm*. Packets are grouped into batches according to their arrival times. The basic idea is to run MWSS on one batch, then take a break to clear the backlog for that batch. After that, the batch is flushed out of the buffers, and the algorithm begins afresh with the next batch. Note that these breaks will cause a loss in throughput, since the MWSS algorithm is now running for only a fraction of the time. However, by choosing a large enough batch length, this throughput loss can be made arbitrarily small. The algorithm is described in more detail below.

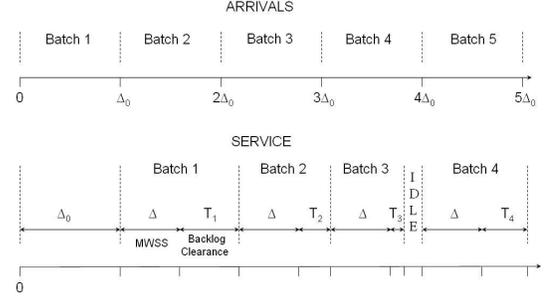

Fig. 3. A typical instance of the batching process in the finite horizon MWSS algorithm

The batch length is denoted $\Delta_0$, and all arrivals from time $k\Delta_0 + 1$ to $(k+1)\Delta_0$ are said to belong to batch $k$, for $k = 0, 1, 2, \ldots$ The processing of a batch begins only after it has fully arrived. Each batch is processed as follows. For a frame of $\Delta(< \Delta_0)$ time-slots, the MWSS algorithm is allowed to run on packets of the current batch. In order to exactly mimic the MWSS algorithm, we impose the constraint that, at the $k^{th}$ slot of the MWSS frame, the weights and coded packets are computed using only those arrivals that occurred before slot number $\frac{\Delta_0}{\Delta}k$ in the batch, even though the entire batch is available. This is because the original MWSS algorithm runs in an online manner, without using future arrivals. This restriction allows us to use the stability result of Theorem 4 for the finite horizon case. It is expected that using the entire batch at every step will only improve the performance.

At the end of the frame, the switch clears out the existing backlog in the degrees of freedom of the current batch, by sending enough number of innovative packets to each of the sub-flows one by one. The duration of this backlog clearance period depends on the amount of backlog that the MWSS algorithm leaves behind. We denote this duration by $T_k$ for the $k^{th}$ batch. Thus, whatever arrives in a time $\Delta_0$ is cleared within a time of $(\Delta + T_k)$.

Once the backlog of the $k^{th}$ batch is cleared, all packets of that batch are flushed out and a new frame begins, during which the next batch is processed. Before starting the new frame, the algorithm waits for the $(k+1)^{th}$ batch to arrive completely at the switch. If there is no fully arrived batch waiting at the switch, and all previous batches have been served, the system is said to be in an *idle state*. All other times are called *busy state*. A typical instance of the batching process is shown in Fig. 3.

*Theorem 5:* If the arrival rate vector is strictly inside $\Gamma$,

*then there exist choices of $\Delta$ and $\Delta_0$ for which, the finite horizon MWSS algorithm guarantees stability of the buffers in the switch, in the sense that the system reaches an idle state infinitely often, w.p. 1.*

*Proof:* We use the stability of the MWSS algorithm in terms of degrees of freedom, to show that for each batch, the expected backlog clearance time can be made as small as needed, compared to the size of the frame $\Delta$.

For any rate point $\mathbf{r}$ that is strictly inside the rate region, $\exists \epsilon > 0$, s.t. $(1+\epsilon)\mathbf{r}$ is also inside the rate region. Choose $\Delta_0 = (1+\epsilon)\Delta$. Consider a single MWSS frame. As far as the algorithm is concerned, the arrival process appears like the original arrival process, except that, the time-axis is compressed by a factor of $\frac{\Delta_0}{\Delta}$. As a result, the MWSS algorithm sees an effective arrival rate of $\frac{\Delta_0}{\Delta}\mathbf{r}$. Since this effective rate is inside the rate region, it follows from Theorem 4 that the backlogs $x_{iJj}(t)$ are stable in the mean.

Now, the backlog clearance duration $T_k$ is essentially the sum of the backlogs in terms of degrees of freedom over all sub-flows at the end of the MWSS frame, *i.e.*, $T_k(\Delta) = \sum_{(i,J,j)} x_{iJj}(\Delta)$. Thus, $T_k$ is also stable in the mean, *i.e.*, $\lim_{t\to\infty} E[T_k(t)] < \infty$. It follows that: $\lim_{t\to\infty} \frac{E[T_k(t)]}{t} = 0$. Choose $\Delta$ large enough such that, $\frac{E[T_k(\Delta)]}{\Delta} < \epsilon$.

To prove the stability of the buffers, we use the notions of idle state and busy state. The time for which the system is in a busy state, is called a *busy period*. We prove that the duration of a single contiguous busy period is finite with probability 1. This implies that the system is stable in the sense that it will become empty infinitely often with probability 1.

Consider a single busy period. We denote the waiting time of the $n^{th}$ batch in the busy period, by $W_n$. This is the difference between the time when the batch arrives completely at the switch and the time when the batch is flushed out after service. In a busy period, $W_n$ is always more than $\Delta_0$. The moment $W_n$ falls below $\Delta_0$, the busy period ends, since any batch arrives only at the end of $\Delta_0$ slots after the previous one. Hence, we need to show that $W_n$ will fall below $\Delta_0$ eventually.

Now,
$$W_n = n\Delta + \sum_{i=1}^{n} T_i - (n-1)\Delta_0.$$

Thus, $W_n < \Delta_0$ iff $\sum_{i=1}^{n} T_i < n\epsilon\Delta$. Since the $T_i$'s are *i.i.d.*, it follows from the strong law of large numbers that for large $n$, *w.p. 1*,
$$\sum_{i=1}^{n} T_i = nE[T_i] + o(n).$$

(Here, a function $f(n)$ is said to be $o(n)$, if $\lim_{n\to\infty} \frac{f(n)}{n} = 0$.)

Now, since $E[T_i] < \epsilon\Delta$, $\sum_{i=1}^{n} T_i$ can be made smaller than $n\epsilon\Delta$, for large enough $n$. This completes the proof. ∎

*Remark 1:* The algorithm used in this proof can be improved using an online streaming policy for buffer clearance, in place of a frame based policy. For instance, packets in the buffer can be replaced with innovative linear combinations in every time-slot. The analysis of this approach is part of future work.

*Remark 2:* The results in [24] are related to our approach. In that paper, the authors analyze the performance of a back-pressure based policy for wired and wireless networks with intra-flow coding, and show that it stabilizes the system for all rate vectors within the capacity region. The network constraints are captured in terms of capacities on each link, which could be inter-dependent in the wireless setting. The crossbar switch, studied in our paper, is similar to the wireless setting in the sense that, an input may not send a different packet to different outputs simultaneously. Besides, there is a special kind of inter-dependence among the links in that, of all links going to the same output, at most one may be active at a time. However, [24] gives an indirect characterization of the rate region in terms of certain flow variables, unlike the more explicit graph-theoretic characterization we have provided.

### A. Simulation of the Online Algorithm

In this section, we study the effect of coding in an online setting, through MATLAB simulations in a $4 \times 3$ switch. The setup we use is similar to the finite horizon MWSS algorithm, except, instead of the max weighted stable set which is known to be $NP$-hard [25], we use a simpler randomized algorithm using the idea proposed in [23]. In each slot, we choose the best of a constant number of randomly generated maximal stable sets, and the stable set that was used in the previous slot. The values of $\Delta$ and $\epsilon$ were chosen to be 1000 slots and 0.005 respectively.

We compare the performance with the case of fanout splitting without coding. For this case, we use a similar randomized modification of the algorithm given in [11] – instead of stable sets, we work with the modified departure vectors defined in that paper.

The traffic pattern is chosen to be a combination of the example pattern used in Section II weighted by a factor of $\frac{2}{3}\alpha$, and a pattern with uniform unicasts, each having a rate of $0.01\alpha$, where $\alpha$ represents the load factor. Thus, the traffic pattern consists of one broadcast from input 1, with a rate of $\frac{4}{9}\alpha$. There are 3 unicasts, one to each output, from inputs 1, 3 and 4, each having a rate of $0.01\alpha$. From input 2, there is a unicast of rate $(\frac{2}{9} + 0.01)\alpha$. The arrivals are generated according to an *i.i.d.* Bernoulli process independently for each flow.

Fig. 4 shows the plot of delay vs. load for the randomized algorithm with and without coding. At light loads, the algorithm due to coding is seen to have a larger delay. However, for the uncoded scheme, the delay shoots up at a lower value of load, as opposed to the coded scheme. Thus, in terms of throughput, the coded scheme is clearly better. Equivalently, we can say that at high loads, network coding leads to benefits in terms of delay.

## VI. CONCLUSIONS

This paper addresses the problem of serving multicast flows in an input-queued crossbar switch. We study the effect of allowing linear intra-flow network coding at the inputs. We show that network coding leads to a larger rate region in general, and demonstrate examples of traffic patterns where coding eliminates the need for speedup, to serve the traffic in a stable manner. We use a graph-theoretic formulation to derive the rate region of the switch with network coding, and propose offline and online algorithms to achieve this rate region. We also use the graph-theoretic formulation to understand the effect of the structure of the traffic pattern on the throughput and on the complexity of computing the schedule. Possible future work could be to use this formulation to come up with approximation schemes and heuristics that simplify the online scheduling algorithm and make it practical.

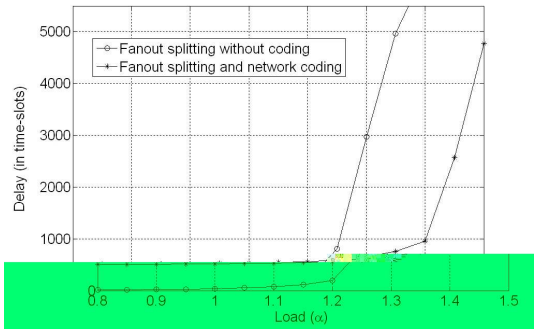

Fig. 4. Delay vs. load plots with and without network coding

In summary, by allowing intra-flow linear network coding, we get not only a gain in throughput, but also a more insightful characterization of the rate region, with potential to use graph-theoretic results and algorithms.